\pdfoutput=1

\documentclass[fleqn,usenatbib]{mnras}

\usepackage{newtxtext,newtxmath}
\usepackage[T1]{fontenc}
\usepackage{ae,aecompl}

\usepackage{graphicx}	% Including figure files
\usepackage{amsmath}	% Advanced maths commands
\usepackage{amssymb}	% Extra maths symbols

\title[The use of CNNs for strong lens modelling]{The use of convolutional neural networks for modelling large optically-selected strong galaxy-lens samples}

\author[J. Pearson, N. Li and S. Dye]{
James Pearson,$^{1}$\thanks{E-mail: james.pearson@nottingham.ac.uk}
Nan Li,$^{1}$
and Simon Dye$^{1}$
\\
$^{1}$School of Physics and Astronomy, Nottingham University, University Park, Nottingham, NG7 2RD, UK\\
}

\date{Accepted XXX. Received YYY; in original form ZZZ}

\pubyear{2019}

\begin{document}
\label{firstpage}
\pagerange{\pageref{firstpage}--\pageref{lastpage}}
\maketitle

\begin{abstract}
We explore the effectiveness of deep learning convolutional neural networks (CNNs) for estimating strong gravitational lens mass model parameters. We have investigated a number of practicalities faced when modelling real image data, such as how network performance depends on the inclusion of lens galaxy light, the addition of colour information and varying signal-to-noise. Our CNN was trained and tested with strong galaxy-galaxy lens images simulated to match the imaging characteristics of the Large Synoptic Survey Telescope (LSST) and Euclid. For images including lens galaxy light, the CNN can recover the lens model parameters with an acceptable accuracy, although a 34 per cent average improvement in accuracy is obtained when lens light is removed. However, the inclusion of colour information can largely compensate for the drop in accuracy resulting from the presence of lens light. While our findings show similar accuracies for single epoch Euclid VIS and LSST $r$-band data sets, we find a 24 per cent increase in accuracy by adding $g$- and $i$-band images to the LSST $r$-band without lens light and a 20 per cent increase with lens light. The best network performance is obtained when it is trained and tested on images where lens light exactly follows the mass, but when orientation and ellipticity of the light is allowed to differ from those of the mass, the network performs most consistently when trained with a moderate amount of scatter in the difference between the mass and light profiles.
\end{abstract}


%%%%%%%%%%%%%%%%%%%%%%%%%%%%%%%%%%%%%%%%%%%%%%%%%%

%%%%%%%%%%%%%%%%% BODY OF PAPER %%%%%%%%%%%%%%%%%%

\section{Introduction}

Strong galaxy-scale gravitational lensing has proven itself to be a highly versatile tool for probing many different physical properties of the Universe. Such lensing can help constrain the mass density profile and hence the dark matter content of the foreground galaxy \citep[e.g.][]{bolton2012boss}, and if combined with redshift measurements this can aid in galaxy evolution models. The observed distortion encapsulates the projected mass density profile, with measurements of such profiles containing information on the dark matter content and substructure within the lens. Advancements have recently been made towards detecting substructure in the lensing mass \citep{vegetti2009bayesian,hezaveh2016detection,ritondale2019low}, which could shed further light on how galaxies evolve. When combined with other methods such as galaxy rotation curves \citep{strigari2013galactic,hashim2014rotation}, projected mass profiles can be used to obtain the amount of dark matter and potentially an approximate 3D (deprojected) mass density profile, such profiles being useful for testing General Relativity \citep{collett2018precise} and cosmological models \citep{eales2015h,krywult2017vimos,davies2018galaxy}. Measurements of gravitational time delays can constrain the value of the Hubble constant irrespective of the distances to or between the galaxies \citep[e.g.][]{bonvin2017h0licow,birrer2019h0licow,chen2018constraining}.

The use of strong lensing for probing high-redshift source populations has also received a recent surge in interest \citep[e.g.][]{dye2018modelling,lemon2018gravitationally,mcgreer2018bright,rubin2018discovery,salmon2018relics,shu2018sdss}. A large amount of effort has been invested in reconstructing the surface brightness distribution of lensed extended sources. Such lensing maintains the surface brightness of sources resulting in distorted yet magnified images, which when paired with redshift measurements can provide valuable information on galaxy evolution. If the mass profile of the lens is well constrained, the original unlensed morphology can be reconstructed, as shown by \cite{warren2003semilinear}, \cite{suyu2006bayesian} and more recently \cite{nightingale2018autolens}, and its properties studied in more detail, for example, their rotation curves \citep{dye2015revealing,geach2018magnified}.

To date there have been several large lens surveys, for example, the Sloan Lens ACS (SLACS) survey \citep{bolton2006sloan}, the CFHTLS Strong Lensing Legacy Survey \citep[SL2S;][]{cabanac2007cfhtls}, the Sloan WFC Edge-on Late-type Lens Survey \citep[SWELLS;][]{treu2011swells}, the BOSS Emission-Line Lens Survey \citep[BELLS;][]{brownstein2011boss} and lenses found in the Dark Energy Survey \citep{dark2005dark}, but so far only a few hundred strong lenses have been found, with most lying at low redshift. This is set to change in the near future; the European Space Agency's Euclid telescope \citep{laureijs2011euclid} is due to launch in 2021 with the primary aim of studying dark matter and dark energy through measurements of the acceleration of the Universe up to a redshift of $z=2$, covering 15,000 deg$^{2}$ over its six year mission. Meanwhile, the ground-based Large Synoptic Survey Telescope \citep[LSST;][]{ivezic2008large} will begin science operations in 2023, with its planned ten year survey covering around 18,000 deg$^{2}$ in six bands ($u$, $g$, $r$, $i$, $z$, $y$), also studying dark matter and dark energy. Together, these are expected to produce billions of galaxy images containing tens of thousands of strong lensing systems \citep{collett2015population}.

To deal with the vast numbers of images, there has been much work on the development of automated methods for rapidly and accurately identifying strong gravitational lens systems. These have included geometrical quantification techniques \citep{seidel2007arcfinder,bom2017neural}, the analysis of colour bands \citep{gavazzi2014ringfinder,maturi2014multi}, spectroscopic analysis \citep{baron2016weirdest,ostrovski2017discovery}, and the Histogram of Oriented Gradients (HOG) feature extraction method \citep{avestruz2017automated}.

Convolutional Neural Networks (CNNs) have also been extensively used in gravitational lens detection \citep{jacobs2017finding,lanusse2017cmu,petrillo2017finding,metcalf2018strong,schaefer2018deep} as these do not require spectroscopic data nor arbitrary geometric measurements. CNNs have become popular in recent years due to their ability to handle large amounts of data, often in the form of images. A pre-trained CNN is capable of classifying thousands of images extremely quickly; they have been shown to be very effective at identifying lenses purely from images, and are able to do so with great efficiency. However, to do this they must be trained on tens of thousands to hundreds of thousands of images. As there is a lack of real images of gravitational lenses, these must be simulated.

Mass modelling of identified lens systems has traditionally seen the application of parametric techniques, where the lens mass model parameters are optimized until an image is produced that best fits the observed image \citep[e.g.][]{warren2003semilinear,vegetti2009bayesian}. While these methods make use of pixellated grids to reconstruct sources, shapelets have also been used for this purpose \citep{birrer2015gravitational,birrer2018lenstronomy}. Such techniques invariably require the time-consuming process of removing lens galaxy light and image masking prior to modelling. In \cite{nightingale2018autolens}, a comprehensive method to circumvent this initial processing by simultaneously fitting both the lens galaxy light and mass profile was presented but at the expense of further slowing the modelling. With the success of CNNs in lens detection, \cite{hezaveh2017fast} demonstrated the first use of CNNs to estimate the mass model parameters of lenses. They trained a combination of four networks to predict parameters of a singular isothermal ellipsoid (SIE) mass model, with \cite{levasseur2017uncertainties} detailing a method of obtaining uncertainties on these predictions. This was recently extended by \cite{morningstar2018analyzing} for application to interferometric observations, and \cite{morningstar2019data} who demonstrated the use of machine learning to additionally reconstruct the background source from CNN-predicted parameters. In these studies, lens light profiles were omitted from the images before training and testing, with training taking multiple days on a GPU machine. They reported an increase in lens modelling speed of several orders of magnitude compared to parametric methods, demonstrating the potential application of CNNs for this purpose.

In this paper, we take two additional steps, investigating the practicalities of using CNNs for lens modelling. First, we assess whether a network can be trained to reliably model the lens mass profile without prior removal of lens light, including an assessment of the impact of assumed mass and light alignment during training. Secondly, we quantify whether gains in accuracy can be achieved by using multiband imaging. We have trained and tested our networks on images simulated to match the imaging characteristics of both LSST and Euclid, with a view to predicting strong lens modelling efficacy for these surveys. Specifically, we have incorporated the Euclid VIS filter and the $g$, $r$ and $i$ bands of the LSST in our simulated images.

The paper is organised as follows: Section \ref{sec:methodology} details the methodology used in simulating the image data sets, and provides an overview of the architecture of our CNN. Section \ref{sec:results} presents the results from the training and testing the CNN, including a comparison between LSST and Euclid data sets, testing data sets with and without lens light, using multiband images, examining the accuracy as a function of stacked images and signal-to-noise, and investigating the impact that the correlation between lens light and mass profiles has on the network. The results are discussed in Section \ref{sec:summary_discussion} along with a conclusion of this work. Throughout this paper we assume a flat $\Lambda$CDM cosmology using the latest Planck results \citep{aghanim2018planck}, with Hubble parameter $h=0.674$ and matter density parameter $\Omega_{\rm m}=0.32$.

\section{Methodology} \label{sec:methodology}

This work has used supervised machine learning, requiring training on a data set of simulated strong lensed images labelled by their lens model parameters. In Section \ref{subsec:lens_sims} next, we describe our procedures for generating the simulated images. Section \ref{subsec:cnn_architecture} then outlines our adopted network architecture.

\subsection{Lensed image simulations} \label{subsec:lens_sims}

For this work we have assumed that all lenses are early-type galaxies, hence the lens mass profile we adopted was the Singular Isothermal Ellipsoid (SIE) model \citep{keeton2001catalog}, commonly used as a good fit for strong lens profiles. For the same reason a S\'ersic profile was used to model the light profile of the source and lens, with the S\'ersic index randomly drawn from a normal distribution with a mean of 4 (i.e. that of a de Vaucouleurs profile). The relative flux in each filter was scaled according to the magnitudes for the lens and source in a chosen band, determined using the fundamental plane relation from \cite{hyde2009luminosity} for lenses and the double-Schechter luminosity function of \cite{kelvin2014galaxy} for sources. Redshifts for the lens and source were selected from uniform distributions with upper limits of $z_{\rm lens}=2$ and $z_{\rm source}=6$, based on trial data sets and work by \cite{collett2015population}. Spectral energy distributions (SEDs) of appropriate ages were selected uniformly from LSST simulated object SEDs \citep[][available from the LSST GitHub repository]{2014SPIE.9150E..14C} and redshifted accordingly. These SEDs use the Bruzual and Charlot models \citep{bruzual2003stellar} with a \cite{chabrier2003galactic} IMF, and either an exponential decline or an instantaneous burst of star formation.

The data sets used for training and testing the CNN contained tens of thousands of single or multiband images (see Section \ref{sec:results} for further details). Multiband images were generated using three filters of the LSST with RGB=($i$, $r$, $g$), using the LSST's CCD filter response function \citep[][available from the LSST GitHub repository]{2014SPIE.9150E..14C}, and likewise the Euclid VIS transmission curve. We adopted the native pixel scale of the LSST \citep[0.2 arcsec pixel$^{-1}$;][]{ivezic2008large,abell2009lsst} and Euclid \citep[0.1 arcsec pixel$^{-1}$;][]{racca2016euclid}, and based on the distribution of known Einstein radii the postage stamp images were fixed at $57 \times 57$ pixels. Typical exposures were used for each visit, with LSST images consisting of two 15s exposures \citep{abell2009lsst} and Euclid VIS images consisting of a single 565s exposure \citep{cropper2012vis}, with AB magnitude zeropoints of 27.09, 28.58, 28.50, 28.34, 27.95, 27.18 for LSST \citep{ivezic2010lsst} and 25.5 for Euclid \citep{collett2015population}.

After generating the lensed image of the background source, light from the lens was added and the image then convolved with a Gaussian point spread function with a full width at half-maximum of 0.7 and 0.17 arcseconds for LSST and Euclid, respectively. The sky background was added (\cite{ivezic2010lsst} for LSST, \cite{euclid2015python} and \cite{collett2015population} for Euclid VIS) along with shot noise, the expected read noise (five electrons per readout) and dark current \citep[two electrons per pixel per second;][]{radeka2009lsst}. For this work, multiple data sets required the removal of the foreground lens light. This was implemented by subtracting the true light profile convolved with the point spread function, leaving only shot noise residuals.

For these postage stamp images the lens centre positions followed a normal distribution about the image centre with a standard deviation of one pixel, and for all galaxies the offset between centre of the mass and centre of the light was taken to be zero. The data sets used for training the network consisted of lenses whose Einstein radii, orientations and axis ratios were uniformly distributed to ensure sufficient training across parameter space.

\begin{figure}
	\includegraphics[scale=0.53]{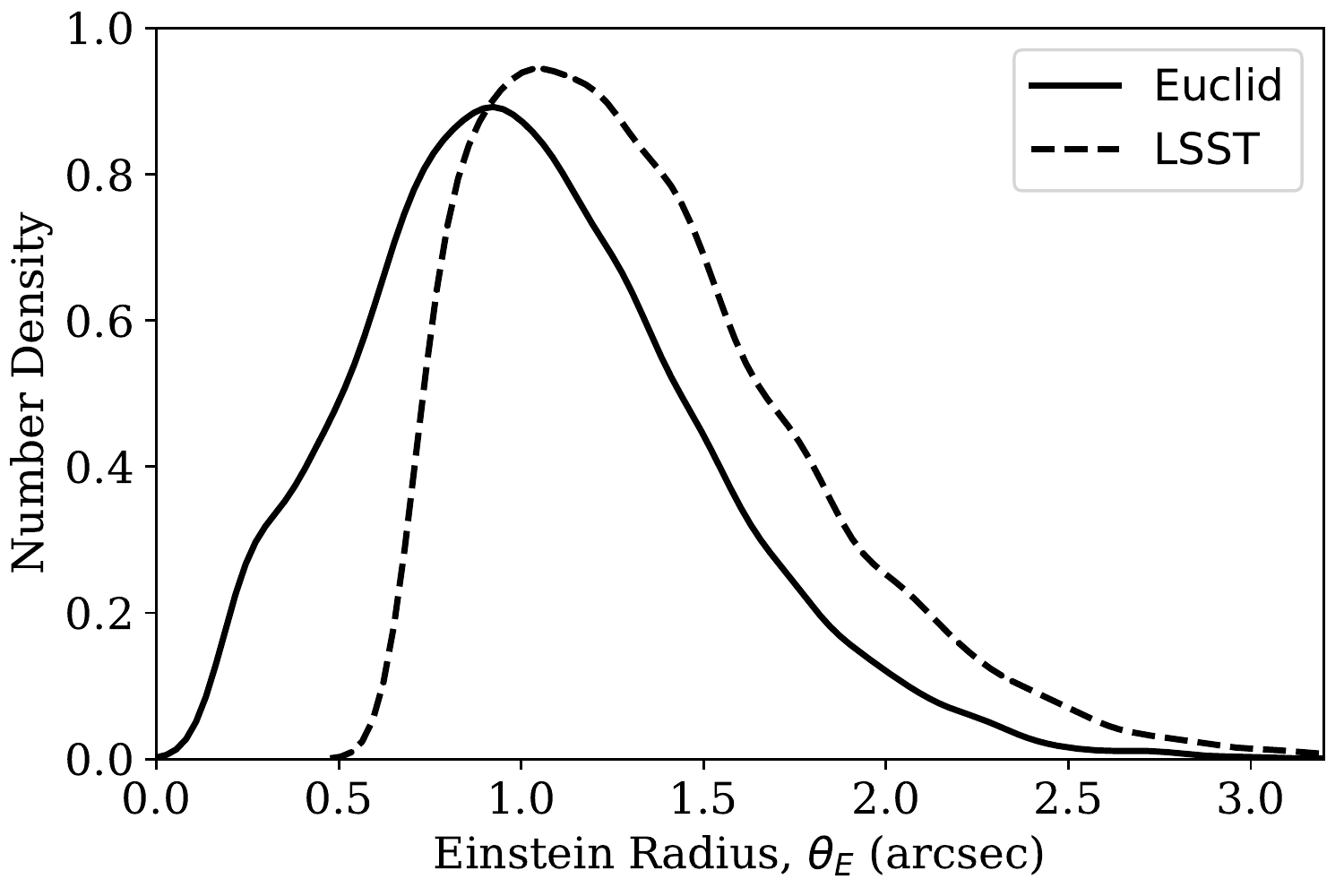}
    \caption{Number density distribution of Einstein radii from the gravitational lens simulator for Euclid and LSST test data sets. When fitted with normal distributions, Euclid has mean 0.97 and width 0.45 arcsec, and LSST has mean 1.22 and width 0.40 arcsec.}
    \label{fig:dist_thetaE}
\end{figure}

\begin{figure}
	\includegraphics[scale=0.53]{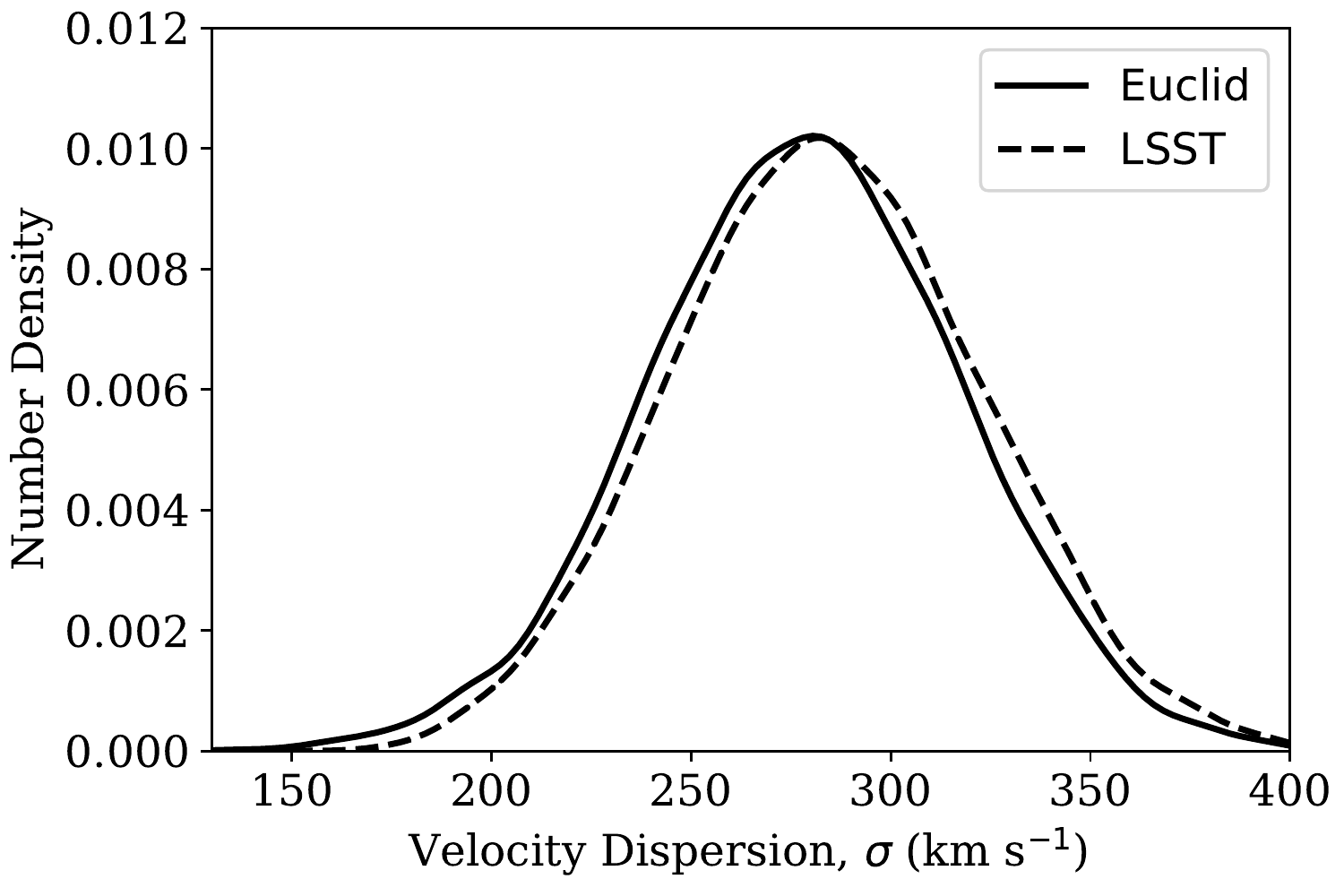}
    \caption{Number density distribution of velocity dispersions from the gravitational lens simulator for Euclid and LSST test data sets. When fitted with normal distributions, Euclid has mean 279 and width 39 km $\rm s^{-1}$, and LSST has mean 283 and width 40 km $\rm s^{-1}$.}
    \label{fig:dist_vel_disp}
\end{figure}

\begin{figure}
	\includegraphics[scale=0.53]{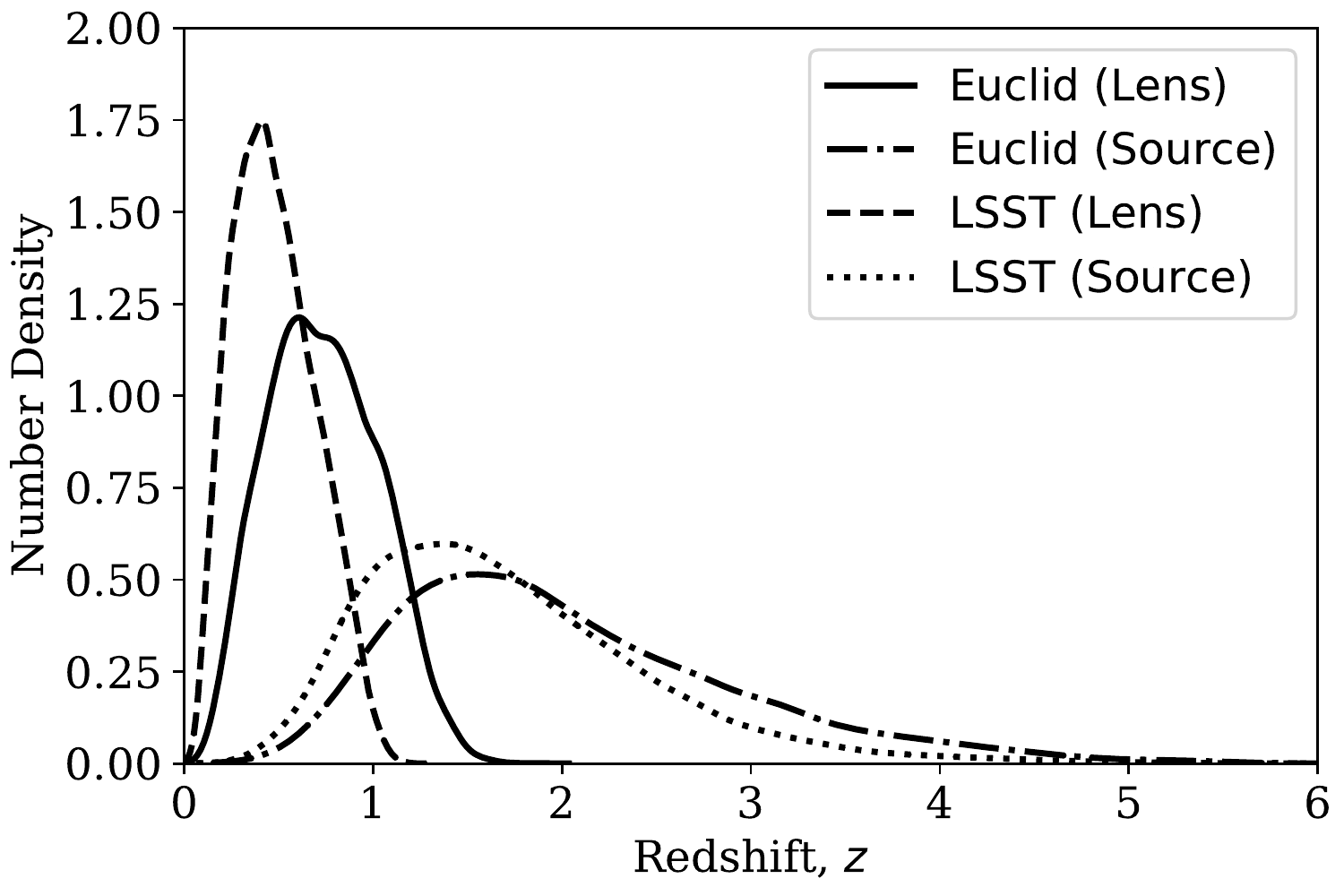}
    \caption{Number density distribution of lens and source redshifts from the gravitational lens simulator for Euclid and LSST test data sets. When fitted with normal distributions, Euclid (lens) has mean 0.72 with width 0.32, Euclid (source) has mean 1.77 with width 0.75, LSST (lens) has mean 0.45 with width 0.23 and LSST (source) has mean 1.51 with width 0.64.}
    \label{fig:dist_redshifts}
\end{figure}

\begin{figure*}
\begin{tabular}{lll}
\includegraphics[scale=3]{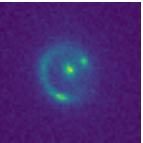}
&
\includegraphics[scale=3]{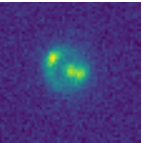}
&
\includegraphics[scale=3]{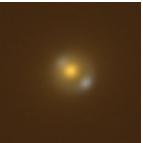}
\end{tabular}
\caption{Example images produced from the gravitational lens simulator. From left to right: simulated Euclid, $r$-band LSST and colour LSST using RGB=($i$, $r$, $g$). Images have a pixel scale of 0.1 arcsec pixel$^{-1}$ for Euclid and 0.2 arcsec pixel$^{-1}$ for LSST.}
\label{fig:typical_sim_images}
\end{figure*}

\begin{figure*}
    \includegraphics[width=17cm]{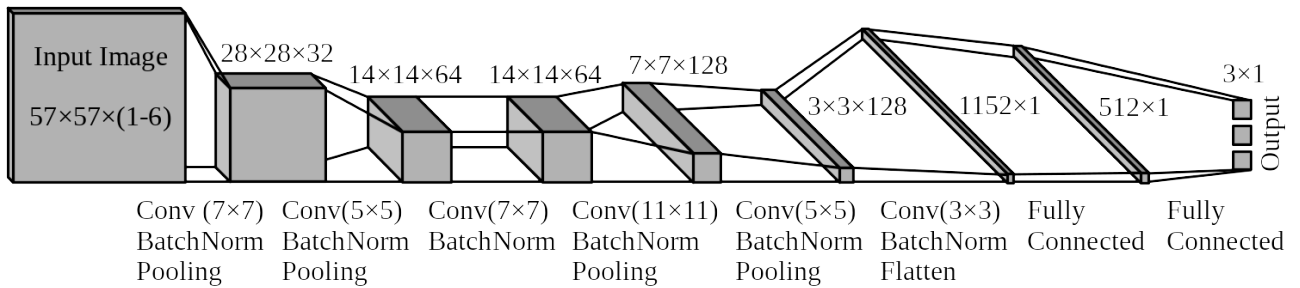}
    \caption{Structure of the CNN used in this work, showing the input image and the output of each block of layers. In total there are six convolutional layers, each with batch normalisation, four max-pooling layers and two fully connected layers. A 'flatten' layer is also included to connect the multidimensional data to the 1D fully connected layer, and ReLU activation is used throughout. Numbers given above or beside each output are the output shapes. The types of layers in each block are given underneath, along with the kernel sizes (in pixels) used. Further details can be found at the end of Section \ref{subsec:cnn_architecture}.}
    \label{fig:cnn_structure}
\end{figure*}

The data sets used for testing the network were simulated to have distributions of parameters in line with the expected observations of LSST and Euclid. Orientation of the light profile was uniformly distributed, while axis ratios $q_{\rm SIE}=b/a$ for the mass profiles were selected from a normal distribution of mean 0.78 and standard deviation of 0.12 \citep{koopmans2006sloan}. Likewise, velocity dispersions were selected from a normal distribution with mean 250 and standard deviation of 44 km s$^{-1}$ based on results from SLACS papers \citep[e.g.][]{shu2017sloan}. These were then used to determine lens Einstein radii which along with effective radii were in agreement with known distributions \citep{koopmans2006sloan,gavazzi2007sloan,bolton2008sloan}. For these generated test data sets, the distributions of Einstein radii, velocity dispersions and redshifts are given in Figs \ref{fig:dist_thetaE}, \ref{fig:dist_vel_disp}, and \ref{fig:dist_redshifts}. 

For both training and test data sets, the ellipticity and orientation of the lens light profile was scattered relative to the mass profile according to the following distributions measured from real lens samples. Light profile axis ratios were set to $q_{\rm SIE}$ divided by a factor taken from a normal distribution of mean 1.02 and standard deviation of 0.12 \citep{bolton2008sloan7}. The mass profile orientation was sampled from a normal distribution centred on the chosen light profile orientation with standard deviation of 10$^{\circ}$ \citep{bolton2008sloan7}.

The lens and source could be offset from one another to produce arcs rather than full rings, with all source centroid positions determined randomly from a uniform distribution within the Einstein radius of the lens in order to ensure a lensed image. Multiple detection criteria were implemented to ensure strong gravitational lenses were produced, based on the criteria in \cite{collett2015population}:
\begin{enumerate}
\item Centre of the source must be multiply-imaged (must lie within the Einstein radius $\theta_{E}$ for this work),
\item Image and counter image must be resolved \mbox{($\theta_{\rm E}^2 > R_{\rm e, source}^2 + seeing^2/2$)} in at least one band,
\item Sufficient magnification ($\mu_{\rm TOT} > 3$) and tangential shearing ($\mu_{\rm TOT} R_{\rm e, source} > seeing$) in at least one band,
\item Sufficient signal-to-noise ratio (SNR > 20) in at least one band.
\end{enumerate}
Some examples of the outputs of the simulation are shown in Fig. \ref{fig:typical_sim_images}, for Euclid, single-band LSST and multiband LSST (RGB = $i$, $r$, $g$).

\subsection{CNN architecture} \label{subsec:cnn_architecture}

Deep neural networks consist of multiple interconnected layers of nodes. The output of a node is the summation of inputs (outputs of the previous layer) multiplied by the strengths of their connections to the node (weights) along with the strength of the node itself (bias), which is then passed through some non-linear activation function. Convolutional Neural Networks (CNNs) are a subset of neural networks which have grid-like layers that apply convolutional filters (kernels) to their inputs in order to extract information, in a similar manner to image processing techniques \citep[e.g.][]{villa2016artificial}.

The network we constructed for mass model parameter estimation contains six convolutional layers, four pooling layers and two fully connected layers, as shown in Fig. \ref{fig:cnn_structure}. The pooling layers are max-pooling, which output the maximum values of each kernel-sized region of the image. The activation function used throughout is the Rectified Linear Unit \citep[ReLU;][]{nair2010rectified}, which acts non-linearly on the nodes such that any negative values are set to zero. There is also a 'flatten' layer that converts the stacks of images into a one-dimensional vector for use in the fully connected layers. Batch normalisation is performed after each convolutional layer, which normalises the output of these convolutional layers to increase the stability of the network.

The network predicts values for the mass model parameters of the SIE lens: the Einstein radius and the two components of complex ellipticity, given as
\begin{equation}
    e_{1}=\frac{1-q^{2}}{1+q^{2}}\cos{2\phi}\,, \\ e_{2}=\frac{1-q^{2}}{1+q^{2}}\sin{2\phi}\,,
    \label{eq:complex_ellip}
\end{equation}
where $q$ and $\phi$ are the axis ratio and orientation of the mass profile, respectively. To improve the network's performance, the inputs are pre-processed; the input images are all intensity scaled so the counts in each pixel lie in the range 0-1, and the training parameters to be predicted are all rescaled to lie in the same ranges as each other, between zero and ten. After training, all parameters are then rescaled back again for calculating accuracy values, and complex ellipticity is converted to conventional ellipticity and orientation of the lens mass profile. 

Convolutional layers have manually-tuneable hyperparameters such as the number and size of the kernels, and through experimentation it was found that six convolutional layers provided the best efficiency (in terms of accuracy and training time) for this work. We chose a stride of one and incorporated zero-padding to maintain the image size through the convolution.

To train the network the image data set is fed into the network in batches. The error on the predictions is determined using the mean squared error (MSE) of the predicted mass model parameters $y_{\rm pred}$ compared to their true values $y_{\rm true}$,
\begin{equation}
    {\rm MSE}=\frac{1}{n}\sum_{i=1}^{n}{(y_{\rm true}-y_{\rm pred})}^2
	\label{eq:mse}
\end{equation}
for $n$ images. For network optimization, we found that the best results were obtained using the Nadam optimizer \citep{dozat2015incorporating}, which is a combination of methods based on the stochastic gradient descent algorithm. The CNN was run on a GPU machine, allowing for much faster processing; training over 150 epochs on 20,000 single-band images takes less than 20 min, and 3-band images in less than 30 min, while testing on such data sets takes only a few seconds.

In terms of the weights and biases, each layer has the following:
\begin{itemize}
\item Convolutional layer: For input height $x_{1}$, width $x_{2}$ and depth (number of bands) $D$, the input is an $(x_{1}, x_{2}, D)$ matrix, with the input to the first layer being the original image of depth $D=1$, or $D=3$ for multiband images. The output is an $(x_{1}, x_{2}, N)$ matrix, where N is the number of kernels applied in that layer. The bias and weights of each kernel can change as the network trains, but remain fixed as the kernel is applied across the image. Each kernel of size $(k_{1}, k_{2})$ has a depth equal to that of the input, and has an associated bias, resulting in a total of $k_{1} \times k_{2} \times D \times N$ weights and $N$ biases per layer. The kernel sizes used are given in Fig. \ref{fig:cnn_structure}.
\item Batch normalisation layer: When used after a convolutional layer, a batch normalisation layer normalises across each feature map of the input using two trainable parameters. Hence for input depth $D$, this results in $2 \times D$ trainable weights.
\item Max-pooling layer: Pooling uses a $2 \times 2$ kernel with a stride of two, hence for input dimensions $(x_{1}, x_{2}, D)$ the output has dimensions $(\rm floor(x_{1}/2), \rm floor(x_{2}/2), D)$.
\item First fully connected layer: The input is a flattened 1152-node array [the preceding layer has output dimensions (3, 3, 128)], and the output is a 512-node array, hence there are $1152 \times 512$ weights and $512$ biases.
\item Final layer: Input is a 512-node array, output is a 3-node array (one output per parameter to estimate), hence there are $512 \times 3$ weights and $3$ biases.
\item Total: 1 band 	= 2,395,267 trainable parameters.
\item Total: 3 bands 	= 2,398,403 trainable parameters. This difference comes from the first layer having deeper kernels for images with multiple bands.
\end{itemize}

\section{Results} \label{sec:results}

Here, we present the results of the investigation, testing the performance of the network on LSST- and Euclid-like data and how performance varies as a function of signal-to-noise. In the following two subsections, we discuss how these results vary between removing and including lens light in the simulated images. In Section \ref{subsec:mass_light_corr} we show how network performance varies with differing assumptions about how mass follows light in both training and test images.

Throughout the remainder of this paper, we quantify the uncertainty on network-predicted lens model parameters by the 68 per cent confidence interval computed from the distributions of differences between true and predicted parameter values across the test image set. We refer to this 68 per cent confidence interval hereafter as a parameter's 'uncertainty'.

\subsection{Removing lens light} \label{subsec:removing_light}

50,000 images with their lens light removed were used to train the CNN to predict the three mass model parameters of the foreground SIE lens: the Einstein radius, and two components of complex ellipticity. After testing the trained CNN on 10,000 images, the predicted components were then converted back to ellipticity (defined as $1-b/a$) and orientation and all results compared to the true values. This process was applied to the simulated LSST $r$-band, LSST multiband (RGB=$i$, $r$, $g$) and Euclid VIS data sets.

\begin{figure}
    \includegraphics[width=\columnwidth]{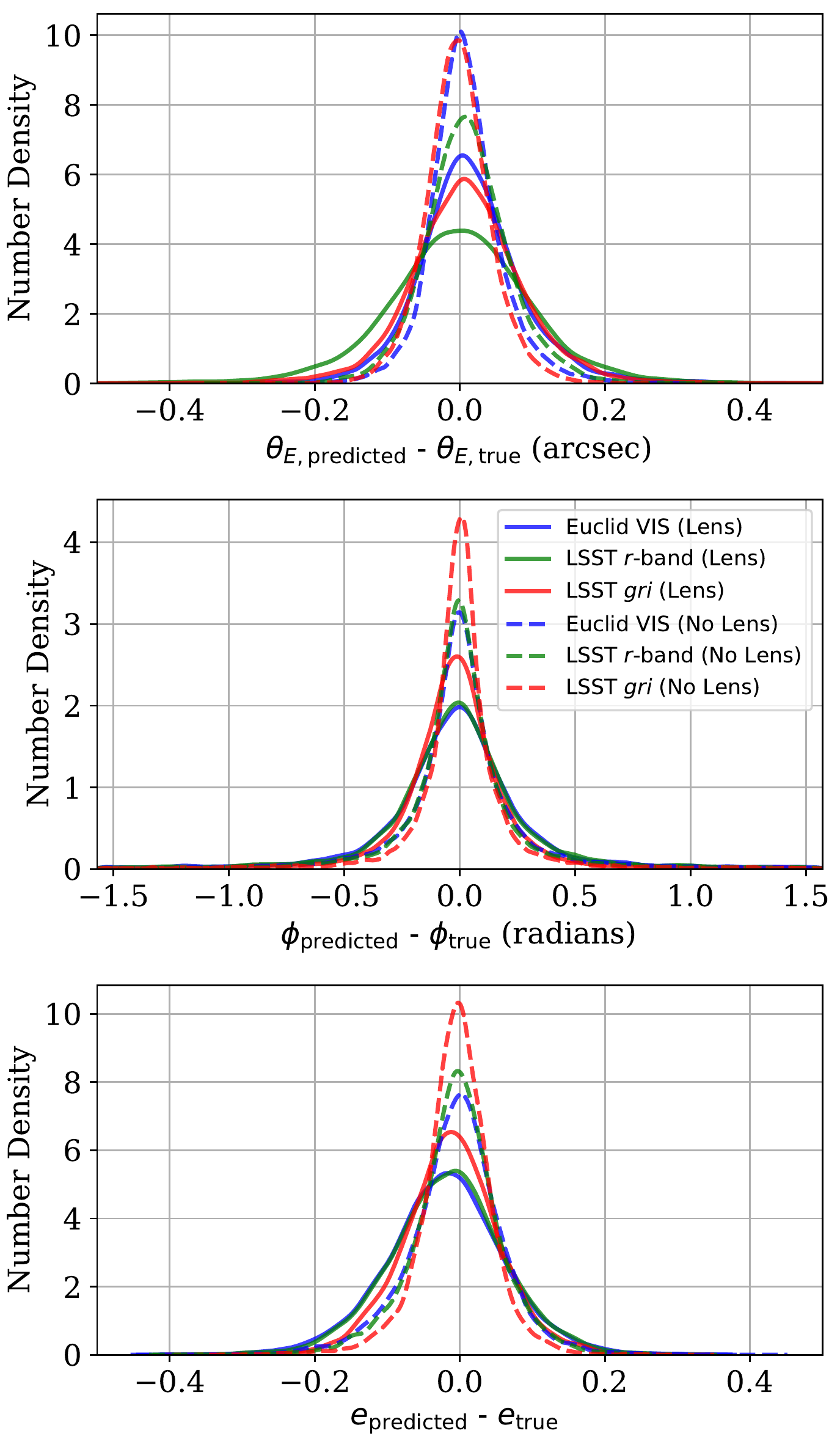}
    \caption{Distribution of the differences between the network's predicted parameters and their true values for test data sets of 10,000 images. From top to bottom: Einstein radius, orientation and ellipticity of the lens mass profile. The distributions shown are those for single-visit Euclid VIS (blue), LSST $r$-band (green) and LSST $gri$ data sets (red), both with the lens light included and removed (solid and dashed lines, respectively).}
    \label{fig:combined_offset_hists}
\end{figure}

\begin{figure*}
    \includegraphics[width=\linewidth]{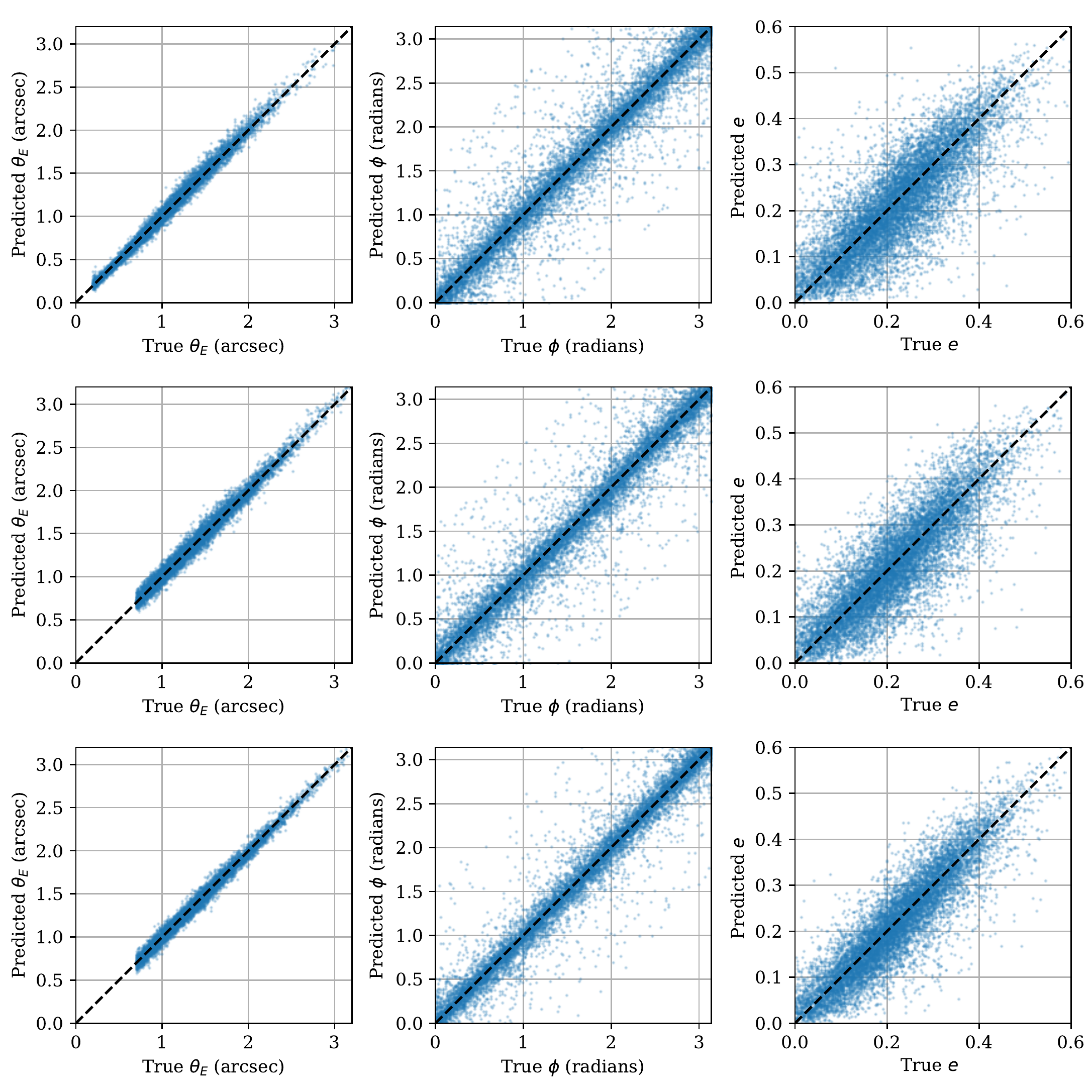}
    \caption{Comparison of network-predicted lens parameters with the true values for images with lens light removed. From left to right: Einstein radius, orientation and ellipticity of the lens mass profile. From top to bottom: Euclid VIS, LSST $r$-band and LSST $gri$.}
    \label{fig:pred_vs_true_nolens}
\end{figure*}

Results are presented in Table \ref{tab:cnn_results_nolens}, given as the uncertainties on the predicted parameters, and in Figs \ref{fig:combined_offset_hists} and \ref{fig:pred_vs_true_nolens}. The results seen in Fig. \ref{fig:pred_vs_true_nolens} show a significantly larger uncertainty for ellipticity compared to the other parameters, as ellipticity has a smaller effect on lens images making it harder to predict.

Training and testing on the single-band LSST data set shows similar results to Euclid for orientation and ellipticity but less accurate estimation of Einstein radius than Euclid owing to lower image resolution and a larger point spread function. The optimal results are those for LSST $gri$, which have the lowest errors for all parameters. The change from using a single LSST band to three bands leads to an average reduction in parameter uncertainty of $24 \pm 2$ per cent (averaged over Einstein radius, orientation and ellipticity). Qualitatively, such an improvement agrees with expectations due to the network being provided with more information.

\begin{table}
	\centering
	\caption{The 68 per cent confidence intervals on network-estimated parameters, computed from the distributions of differences between true and predicted parameter values across 10,000 test images. Results are presented both for images which have had the foreground lens light removed and for images with the lens light present.}
	\label{tab:cnn_results_nolens}
	\begin{tabular}{llll}
		 & Euclid & LSST & LSST\\
        Mass model parameter & VIS & $r$-band & ($gri$)\\
		\hline
		{\bf Images with lens light removed:}\\
		\hline
		Einstein radius (arcsec) & 0.043 & 0.054 & 0.042\\
		$e_{1}$ & 0.066 & 0.062 & 0.047\\
		$e_{2}$ & 0.066 & 0.062 & 0.048\\
		Ellipticity & 0.057 & 0.053 & 0.041\\
		Orientation (radians) & 0.164 & 0.155 & 0.114\\
		\hline
		{\bf Images with lens light included:}\\
		\hline
		Einstein radius (arcsec) & 0.064 & 0.091 & 0.071\\
		$e_{1}$ & 0.089 & 0.090 & 0.072\\
		$e_{2}$ & 0.092 & 0.089 & 0.072\\
		Ellipticity & 0.079 & 0.077 & 0.064\\
		Orientation (radians) & 0.225 & 0.220 & 0.172\\
	\end{tabular}
\end{table}

\begin{figure*}
    \includegraphics[width=\linewidth]{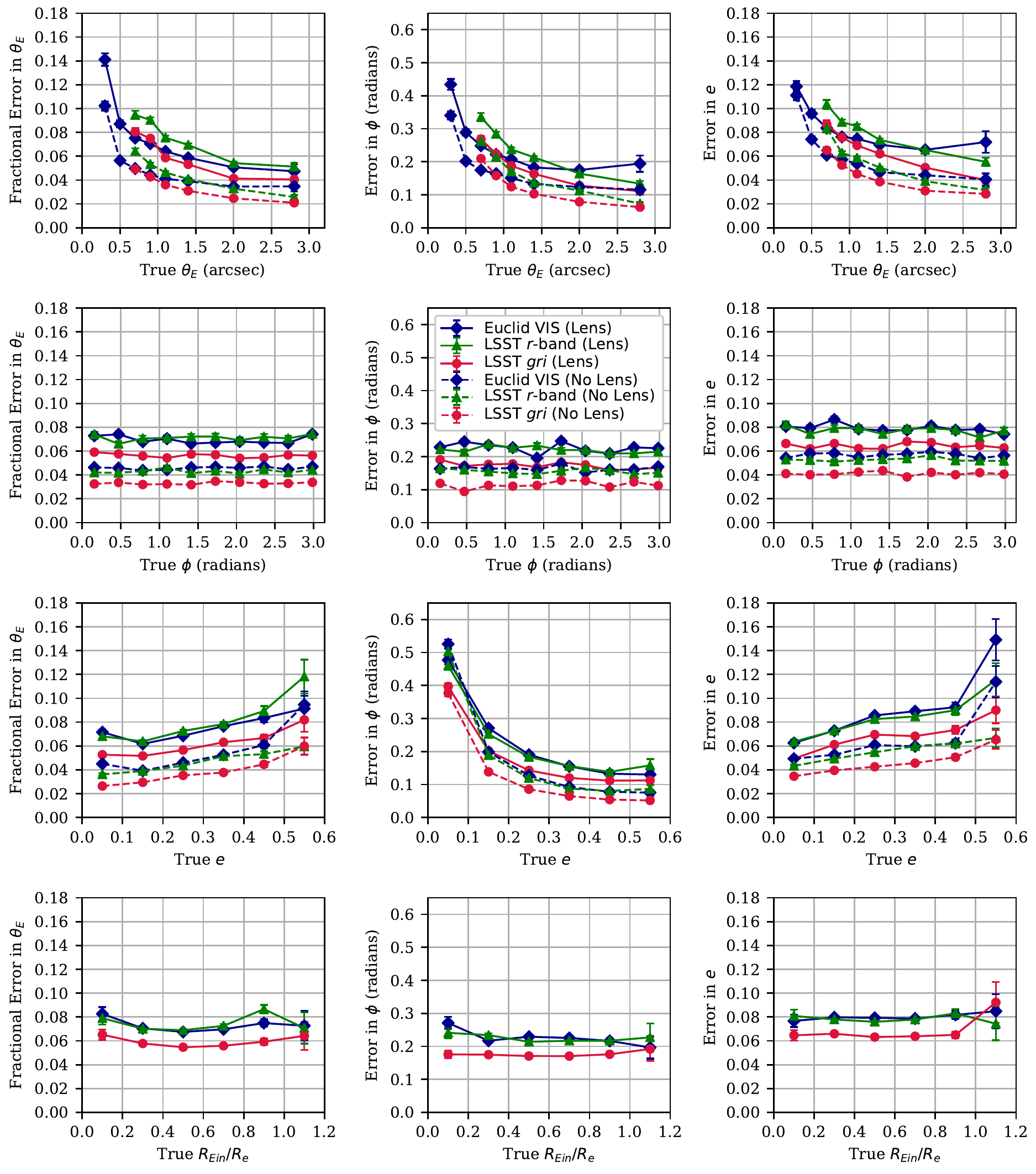}
    \caption{The variation of network-predicted lens model parameter uncertainties with parameter value. The solid and dashed lines correspond to images with lens light included and removed, respectively. The blue diamonds, green triangles, and red circles correspond to Euclid VIS, LSST $r$-band and LSST $gri$ images, respectively. From left to right: Einstein radius, orientation and ellipticity of the lens mass profile. From top to bottom: Einstein radius, orientation, ellipticity and $R_{\rm Ein}/R_{\rm e}$, the ratio between Einstein radius and effective radius. The error in Einstein radius is given as the fractional error, and the error bars are the corresponding standard errors.}
    \label{fig:error_vs_parameters}
\end{figure*}

\begin{figure*}
    \includegraphics[width=\linewidth]{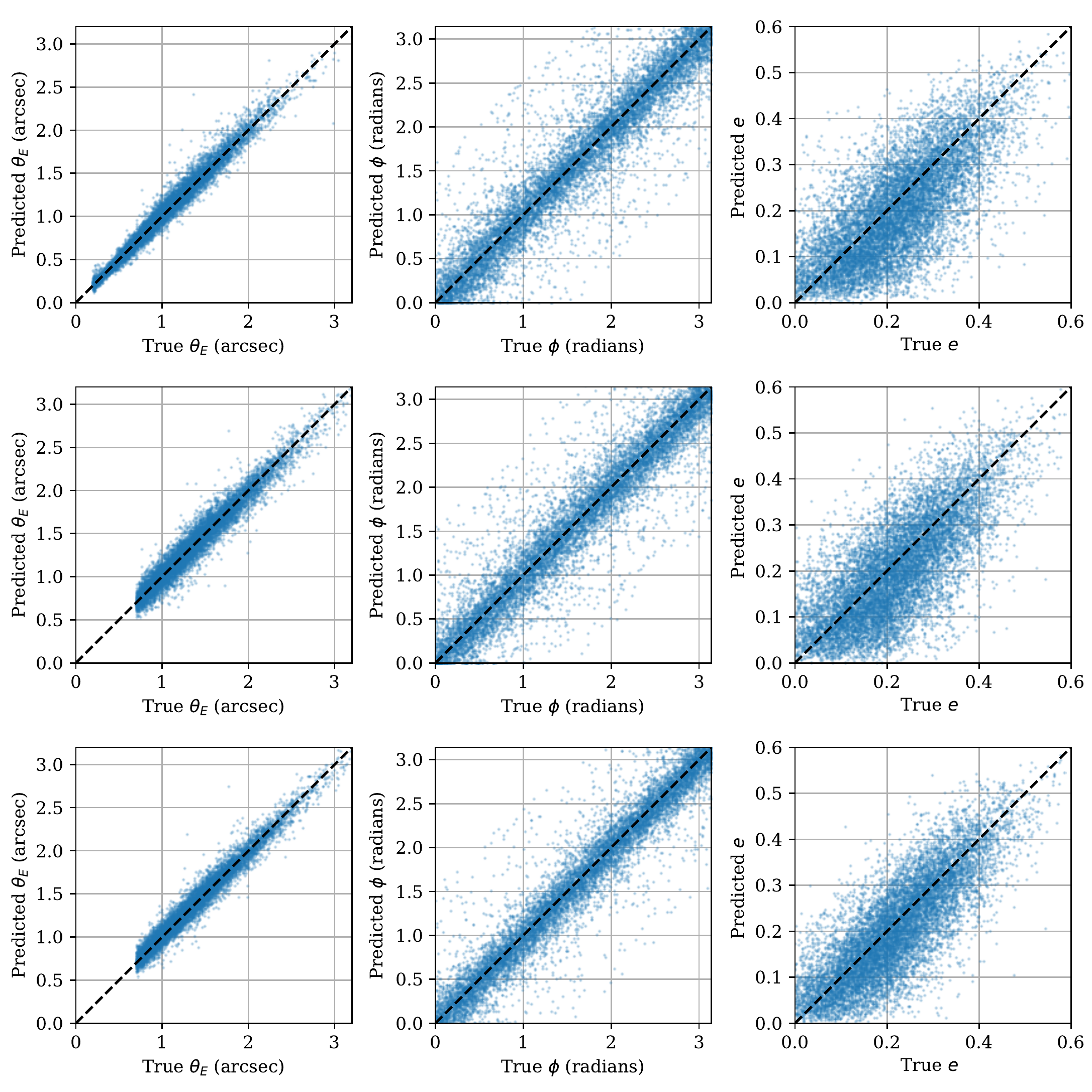}
    \caption{Comparison of network-predicted lens parameters with the true values for images including lens light. From left to right: Einstein radius, orientation and ellipticity of the lens mass profile. From top to bottom: Euclid VIS, LSST $r$-band and LSST $gri$.}
    \label{fig:pred_vs_true}
\end{figure*}

Test results are also shown in Fig. \ref{fig:error_vs_parameters}, presented as the uncertainty for each parameter as a function of the value of each parameter. The results for (angular) Einstein radius $\theta_{E}$ are given as fractional errors, (i.e. 68 per cent confidence interval of $(\theta_{\rm E}^{\rm predicted}-\theta_{\rm E}^{\rm true})/\theta_{\rm E}^{\rm true}$), with the error bars in the figure corresponding to the standard errors. The top row of the figure shows that larger errors occur at small values of $\theta_{\rm E}$, due to the limited resolution of the images resulting in very few pixels to characterise smaller Einstein radii. The errors for all parameters decrease with increasing Einstein radius, as expected due to the network having more information on the shapes of larger arcs. The second row shows that the network's predictions of parameters are unaffected by the orientation of the lens, clearly expected as the network should not show a preference for certain orientations.

For the third row, larger errors occur in Einstein radius for more elliptical lenses due to the resulting lensed arcs less closely following the shape of the Einstein ring. The central graph shows the expected decrease in orientation error as a function of ellipticity, as the more elongated the lens the easier it is to identify the direction of this elongation. Finally, the increasing ellipticity error as a function of ellipticity can be explained as higher ellipticities result in larger caustics within which the source galaxy can reside, and hence produce more varied lensed arcs. As such, the trend vanishes when presented as the fractional error in ellipticity.

\subsection{Including lens light}

We repeated the analysis outlined in Section \ref{subsec:removing_light} but using images containing lens light. For both training and testing of the network, we used images where the ellipticity and orientation of the mass profile were scattered about those of the light profile with standard deviations of 0.12 and 10$^{\circ}$, respectively, motivated by observations \citep{koopmans2006sloan,bolton2008sloan7}. For more information, see Section \ref{subsec:mass_light_corr}.

The test results for parameter uncertainties as functions of parameters are again shown in Fig. \ref{fig:error_vs_parameters}, and closely follow the same trends as images without lens light, although higher on average. Additionally, the bottom row presents the uncertainties as functions of the ratio between the Einstein radius and effective radius, $R_{\rm Ein}/R_{\rm e}$, showing that the network is mostly unaffected by this ratio, at least over the range tested. This is an encouraging indication that lens model parameters can still be reliably estimated even when lens light begins to dominate the lensed source image.

The test results for predicted parameters against their true values are shown in Table \ref{tab:cnn_results_nolens} and in Figs \ref{fig:combined_offset_hists} and \ref{fig:pred_vs_true}. As for images without lens light, the figures show that the network obtains smaller errors for Einstein radius using the Euclid-style data set, and that the smallest errors for the other parameters are again achieved using the three-band LSST data set. With the lens light, the change from using a single LSST band to three bands leads to the uncertainty decreasing by $20 \pm 2$ per cent (again averaged over Einstein radius, orientation and ellipticity). As in the previous section, this is due to the network being given additional information, and with the lens light present the colour information can be used to distinguish between the foreground lens and background source. Even if the colouration of the two were similar, the multiple bands still act to improve the signal-to-noise.

Comparing the results with those in Section \ref{subsec:removing_light}, it is clear that the inclusion of lens light degrades the results for all data sets, including the multiband LSST results. The error in predicting orientation changes least since this is dominated by the larger errors associated with predicting orientation for rounder lenses, seen in Fig. \ref{fig:error_vs_parameters}. The removal of lens light from Euclid VIS and LSST $r$-band images achieves errors comparable to or slightly lower than that of LSST $gri$ with lens light, with notably lower Einstein radius error but larger orientation error. With regards to ellipticity and orientation, this is likely due to the use of three bands almost providing sufficient information to balance out any overuse of the lens light profile (see Section \ref{subsec:mass_light_corr}), while for Einstein radius the presence of the lens light makes it harder for the network to accurately identify the lensed source regardless of the number of colour bands.

The lens light removal from three-band LSST images leads to uncertainties decreasing by $41$, $36$ and $34$ per cent for Einstein radius, ellipticity and orientation, respectively. This is compared to a decrease of $34$, $28$ and $27$ per cent for Euclid VIS and $40$, $32$ and $30$ per cent for LSST $r$-band, resulting in an average decrease of $34\pm5$ per cent across the parameters and data sets. Overall, the network achieves the highest accuracy for LSST $gri$ images in which lens light has been removed, having the lowest errors for all parameters. Clearly networks trained on images without lens light still benefit significantly from the use of multiple colour bands, which increases the accuracy of all predicted parameters.

\subsection{Testing the correlation between light and mass profiles} \label{subsec:mass_light_corr}

We have shown that the presence of lens light is a factor that can influence network accuracy. This conclusion was reached by training and testing with lens images where ellipticities and orientations could differ between the mass and light profiles. Such a scatter is observed in the real Universe \citep[e.g.][]{koopmans2006sloan,bolton2008sloan7}, and can impact the offsets of lensed sources \citep{harvey2016systematic}. In this subsection, we report on a series of tests conducted to assess how well the network performs with varying mismatch between mass-light alignment in the training and testing images. We controlled the mismatch by drawing the ellipticity and orientation of the light profile from two normal distributions with means set to the corresponding values of the already selected mass profile (as described in Section \ref{subsec:cnn_architecture}) and variable standard deviations, such that
\begin{equation}
    \phi_{\rm light}=\mathcal{N}(\mu=\phi_{\rm SIE},\,\sigma_{\phi})\,,
	\label{eq:mass_light_corr_orientation}
\end{equation}
\begin{equation}
    e_{\rm light}=1-q_{\rm light}=1-\frac{q_{\rm SIE}}{\mathcal{N}(\mu=1,\,\sigma_{e})}\,,
	\label{eq:mass_light_corr_ellipticity}
\end{equation}
where $q=b/a$ is the axis ratio, $\mathcal{N}$($\mu$,$\sigma$) is a normal distribution of mean $\mu$ and standard deviation $\sigma$ and the standard deviations of the orientation offset and ellipticity offset are $\sigma_{\phi}$ and $\sigma_{e}$, respectively. Larger values of $\sigma_{e}$ and $\sigma_{\phi}$ correspond to less correlation between the light and mass profiles. For the first two tests, the $\sigma$ for one parameter was varied in the training and test data while the other was kept at $\sigma=0$ in the two data sets. For the third test both $\sigma_{e}$ and $\sigma_{\phi}$ were varied by the same multiples of their SLACS values in the training and test data. For these tests, source positions were restricted to within the lens caustic, rather than within the Einstein radius as before, resulting in more highly magnified sources.

\begin{figure}
    \includegraphics[width=\columnwidth]{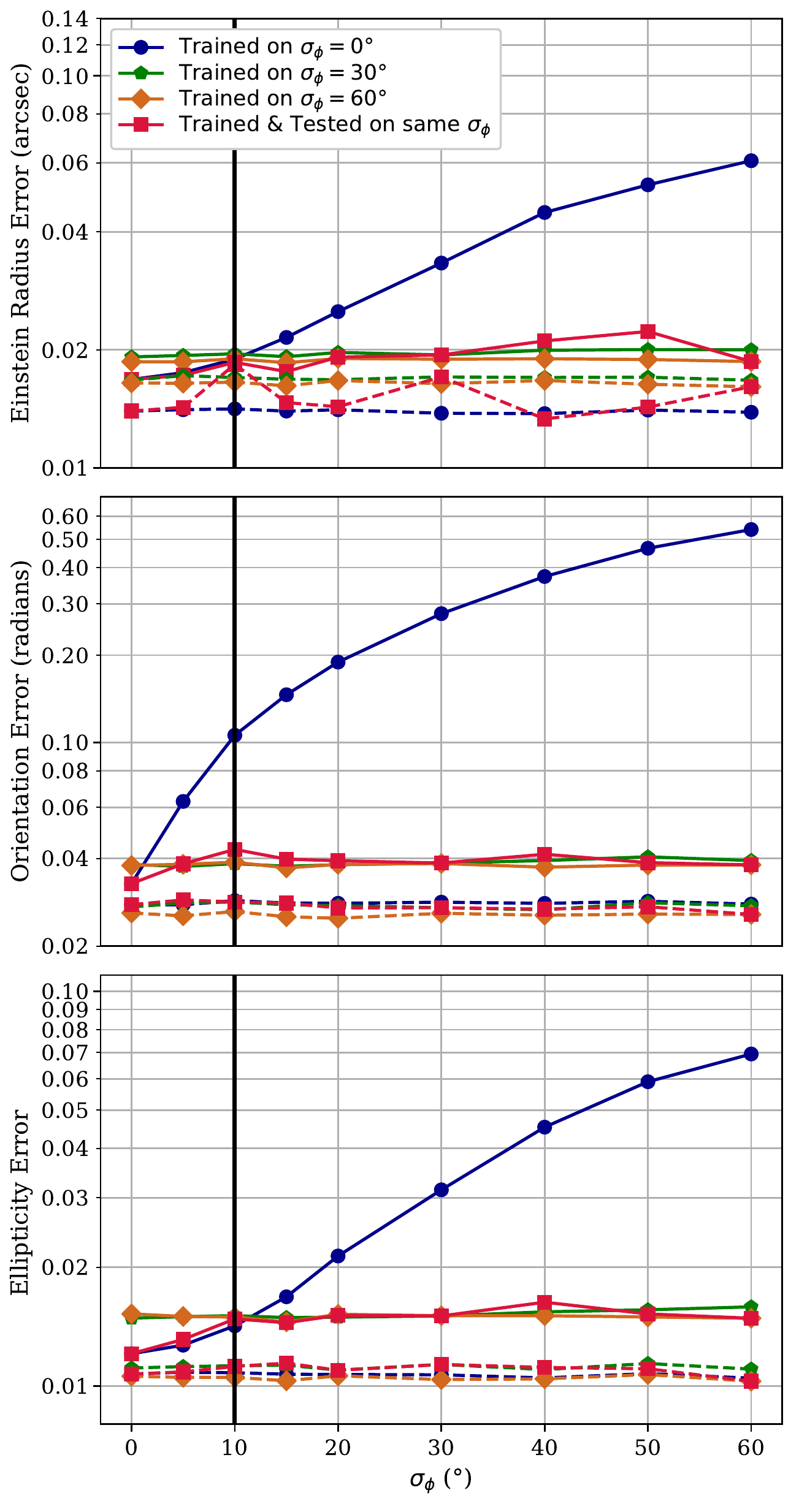}
    \caption{Uncertainties on mass model parameters as a function of the scatter between light and mass orientation in the test data. For all results in this figure, the ellipticity of the light profile follows that of the mass with zero scatter, for both training and test data. From top to bottom: Results for Einstein radius, orientation and ellipticity. The dashed lines indicate the results for data sets with lens light removed, which hence act as a control. The vertical black line indicates the standard deviation used elsewhere in this work, $\sigma_{\phi}=10^{\circ}$, based on SLACS results.}
    \label{fig:error_vs_scatter_orient}
\end{figure}

\begin{figure}
    \includegraphics[width=\columnwidth]{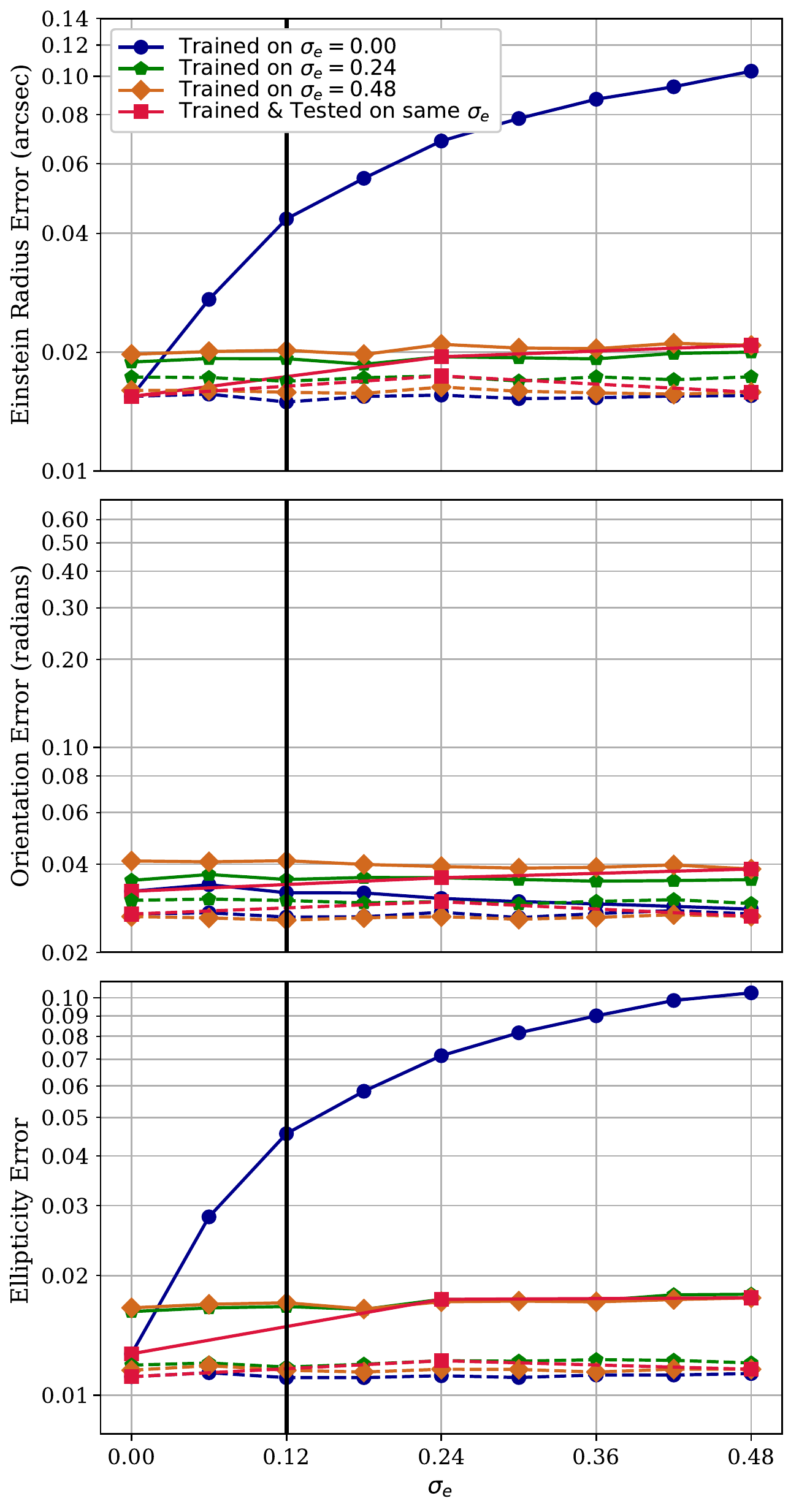}
    \caption{Uncertainties on mass model parameters as a function of the scatter between light and mass ellipticity in the test data. For all results in this figure, the orientation of the light profile follows that of the mass with zero scatter, for both training and test data. From top to bottom: Results for Einstein radius, orientation and ellipticity. The dashed lines indicate the results for data sets with lens light removed, which hence act as a control. The vertical black line indicates the standard deviation used elsewhere in this work, $\sigma_{e}=0.12$, based on SLACS results.}
    \label{fig:error_vs_scatter_ellip}
\end{figure}

\begin{figure}
    \includegraphics[width=\columnwidth]{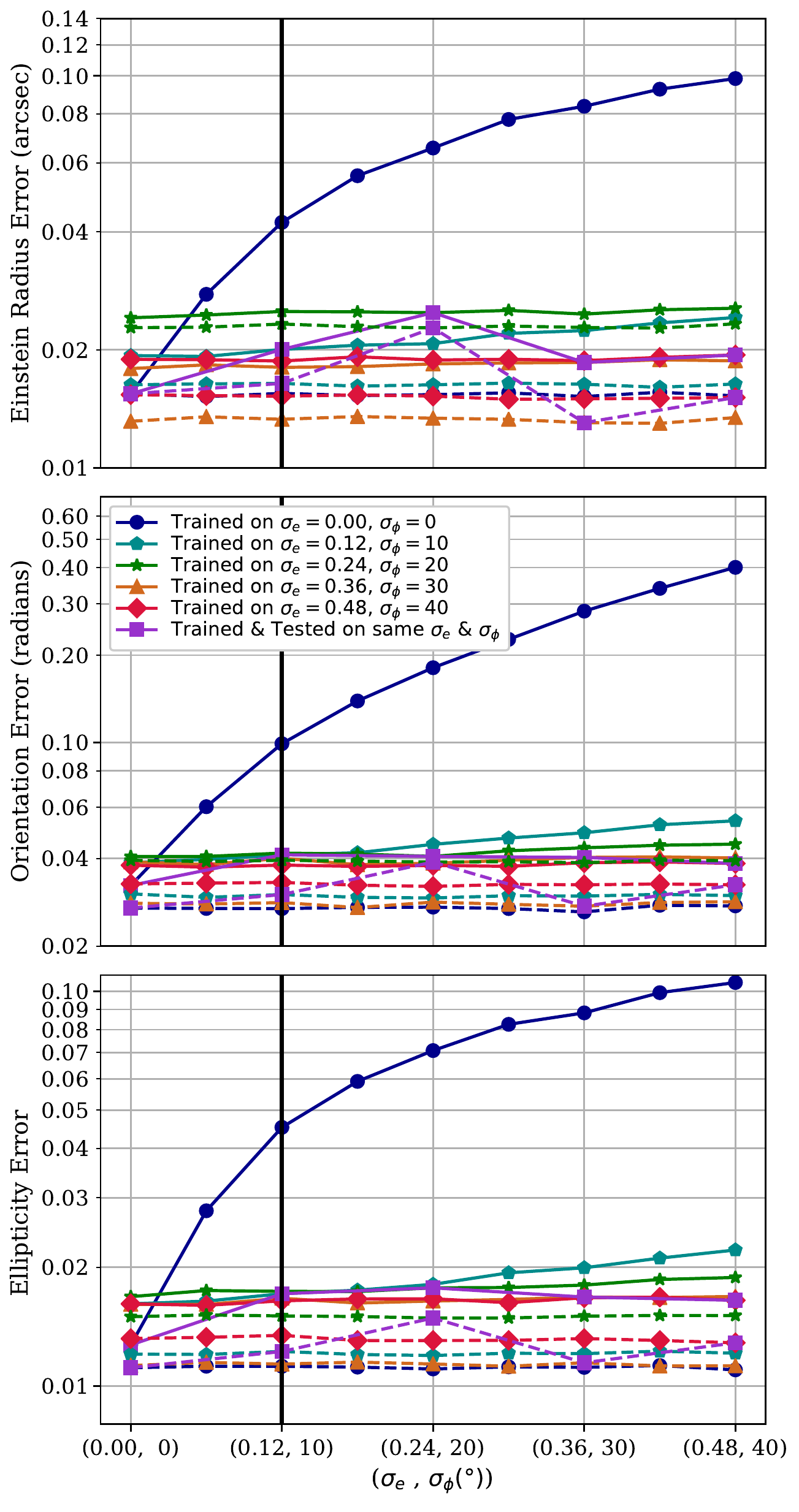}
    \caption{Uncertainties on mass model parameters as a function of the light-mass scatter of both ellipticity and orientation in the test data. From top to bottom: Results for Einstein radius, orientation and ellipticity. The dashed lines indicate the results for data sets with lens light removed, which hence act as a control. The vertical black line indicates the standard deviations used elsewhere in this work, ($\sigma_{e}=0.12$, $\sigma_{\phi}=10^{\circ}$), based on SLACS results.}
    \label{fig:error_vs_scatter_both}
\end{figure}

Notable trends were observed for Einstein radius, orientation and ellipticity errors as functions of $\sigma_{\phi}$, shown in Fig. \ref{fig:error_vs_scatter_orient}. Generally the CNN finds it easier to determine lens parameters when the light profile is orientated similar to the mass profile. However, the Einstein radius remains relatively insensitive to the offset between mass and light orientation regardless of how it is trained, apart from the case when it is trained on images with no offset. The graphs show how a network trained on a scatter of $\sigma_{\phi}$ = 60$^{\circ}$, akin to almost no correlation between the profiles, will vary little with the scatter of the test data set, as it is clearly not associating the two profiles. However, as the scatter between the profiles in the training data set is reduced, the more the network relies on the foreground lens light in determining the orientation of the lens, which can clearly lead to large errors when the network is applied to misaligned systems.

For ellipticity error, while $\sigma_{\phi}$ is varied, the ellipticity of the light and mass profiles are kept identical, so if the network only learned from the lens light then it should perform equally well regardless of $\sigma_{\phi}$. The fact that this is not observed in the lower plot of Fig. \ref{fig:error_vs_scatter_orient} thus indicates that the network is still learning from the lensed source light to an extent when predicting mass ellipticity. Einstein radii are obtained from studying the lensed arcs of the source, and for elliptical lenses different lens orientations and ellipticities for the same source position will result in different arc configurations. When trained on lower $\sigma_{\phi}$ the network associates the lens light with these parameters, and so its predictions of Einstein radii are also less reliable when tested with greater scatter in mass-light orientation. The trend seen is less pronounced as to obtain the Einstein radius the network must still rely more on the lensed source light.

Similar results are also seen in Fig. \ref{fig:error_vs_scatter_ellip} for the network's predictions of Einstein radius and ellipticity as functions of the scatter between light and mass ellipticity in the test data, again with a far weaker trend observed for Einstein radius error. However, for orientation error, larger ellipticity scatter $\sigma_{e}$ in the test data can result in lower prediction errors as shown in the middle panel of Fig. \ref{fig:error_vs_scatter_ellip}. This is because the mass axis ratios of the test data sets are sampled from a normal distribution with a mean of 0.78, and as a result the tail of the distribution is cut off at values of 1 (an ellipticity of zero). This biases the sampling of light profile ellipticities to slightly higher values on average for larger $\sigma_{e}$, and these larger values allow for more accurate orientation predictions, as per Fig. \ref{fig:error_vs_parameters}. The effect on the error is still relatively minor compared to other trends, and an additional explanation for why the error does not increase is that, as discussed when varying $\sigma_{\phi}$, even with the presence of lens light, the network still utilizes the lensed source light to measure ellipticity, and so can use the direction of this distortion to measure orientation.

Fig. \ref{fig:error_vs_scatter_both} shows the results of simultaneously scattering ellipticity and orientation in the test data. With the inclusion of lens light, the lowest average uncertainty is achieved for lenses with zero scatter between light and mass profiles, with values for \{$\theta_{E}$(arcsec), $\phi$(rad), $e$\} of \{0.015, 0.032, 0.013\}. However, this is still an average of $11\pm8$ per cent larger than that of images without lens light, for which errors of \{0.015, 0.027, 0.011\} were obtained by the network. The predicted error is slightly larger than in the previous cases where one of orientation or ellipticity were not scattered, but the trends are similar; training with no scatter results in high prediction errors when mass does not closely follow light whereas training with a moderate amount of scatter, i.e. somewhere between a ($\sigma_{e}$, $\sigma_{\phi}$) of (0.12, 10) and (0.24, 20), gives consistent network performance across a wide range of mismatch between the mass and light. For all cases, except when training on zero scatter, no systematic biases in predicted parameter values were identified. If trained on zero scatter, the network tends to over-estimate the Einstein radius and ellipticity by a small but significant amount of up to 2 per cent.

From these results, the data sets chosen for the rest of this work contained some scatter between the light and mass profiles, with $\sigma_{e}=0.12$ and $\sigma_{\phi}=10^{\circ}$, to both align with observed SLACS distributions and to ensure the network did not rely too heavily on the lens light.

\subsection{Network error as a function of signal-to-noise}

In this section, we show how network accuracy depends on the signal-to-noise ratio (SNR) of the images used for training and testing. We limit our discussion here to images with lens light removed since the previous trends discussed when comparing inclusion and removal of lens light remain the same in this case. Two tests were conducted, with the first investigating how the accuracy of the retrieved lens parameters varies as a function of SNR for both LSST and Euclid images. For the second test, we put this into the context of the Euclid survey and show how retrieved lens parameter accuracy depends on the number of survey visits and how this changes with the number of survey visits assumed during network training.
 
We define SNR as the summed flux $S$ of the lensed source image over the error on this summed flux,
\begin{equation}
    {\rm SNR} = \frac{S}{\sigma \sqrt{N}}
    \label{eq:snr}
\end{equation}
where $\sigma$ is the standard deviation of the background noise and $N$ is the number of pixels making up the signal. For our first test, we kept $\sigma$ fixed and generated a training and test image set, scattering the mass-light orientation and ellipticity offsets with $\sigma_{\phi}=10^{\circ}$ and $\sigma_{e}=0.12$ as discussed in Section \ref{subsec:mass_light_corr}. The randomly drawn lens model parameters give rise to a distribution in SNR within the training and test data. To assess how the network accuracy depends on SNR, we binned the test images by SNR and measured the uncertainty of predicted lens model parameters in each bin.

\begin{figure}
    \includegraphics[width=\columnwidth]{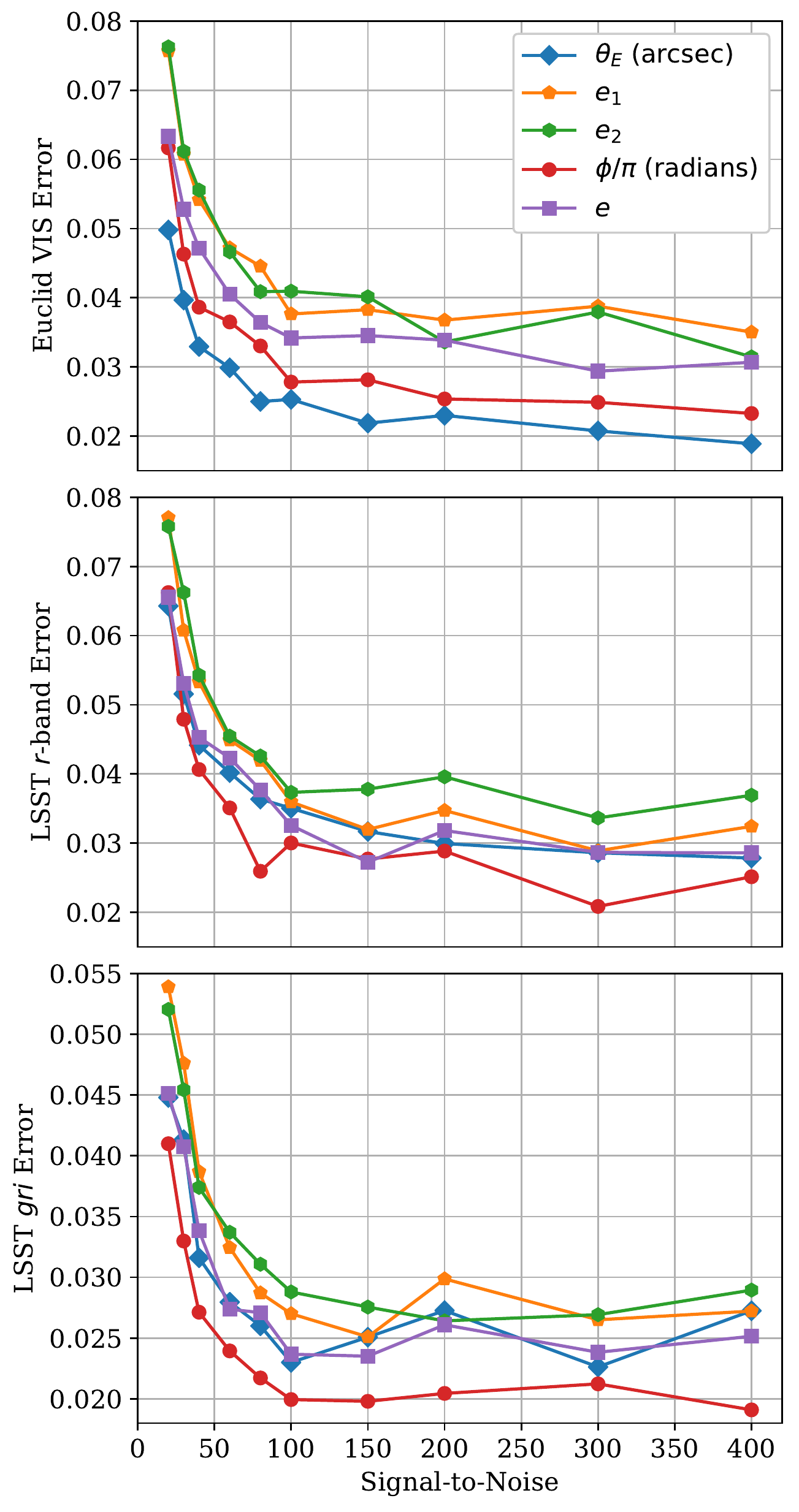}
    \caption{Lens model parameter uncertainties for images with lens light removed, as a function of the signal-to-noise ratio (SNR) of the source in the test images. From top to bottom: simulated images for the Euclid telescope VIS filter, LSST $r$-band and LSST $gri$. Data points lie at the lower SNR bin boundaries, with the last point encompassing all images with SNR$\geq$400. The blue diamonds, orange pentagons, green hexagons, red circles and purple squares correspond respectively to Einstein Radius, the first and second components of complex ellipticity, orientation and ellipticity of the lens. The values of orientation $\phi$ are divided by pi to more easily compare the parameters.}
    \label{fig:error_vs_snr_subplot}
\end{figure}

Fig. \ref{fig:error_vs_snr_subplot} shows for Euclid VIS, LSST $r$-band and LSST $gri$ images, respectively, how the uncertainty decreases with higher signal-to-noise, with little decrease in error past approximately SNR$ = $100. For the highest signal-to-noise images the error can occasionally begin to rise again, due to the network having fewer of these brighter, lower redshift galaxies to train on. These results quantify how the network performs significantly better on higher signal-to-noise images, with the uncertainty of SNR$\geq$400 images a factor of 1.6-2.7 (mean of $2.2\pm0.3$) times smaller compared to that of SNR=20-30 images. This shows how parameter predictions for higher SNR images are much more likely to be closer to their true values, with little change in accuracy beyond around SNR>100.

In our second test, we investigated how the SNR of the training data set affects the performance of the CNN. We produced Euclid image training sets for different values of $\sigma$ in equation (\ref{eq:snr}) corresponding to one (what we have assumed up to now), four \citep[the number of visits expected to be undertaken by Euclid;][]{laureijs2011euclid,cropper2012vis,racca2016euclid}, 10 and 100 visits. The four networks that resulted from training on these data sets were then applied to test images generated with the same range of noise characteristics. All combinations of test and training SNR were investigated.

\begin{figure}
    \includegraphics[width=\columnwidth]{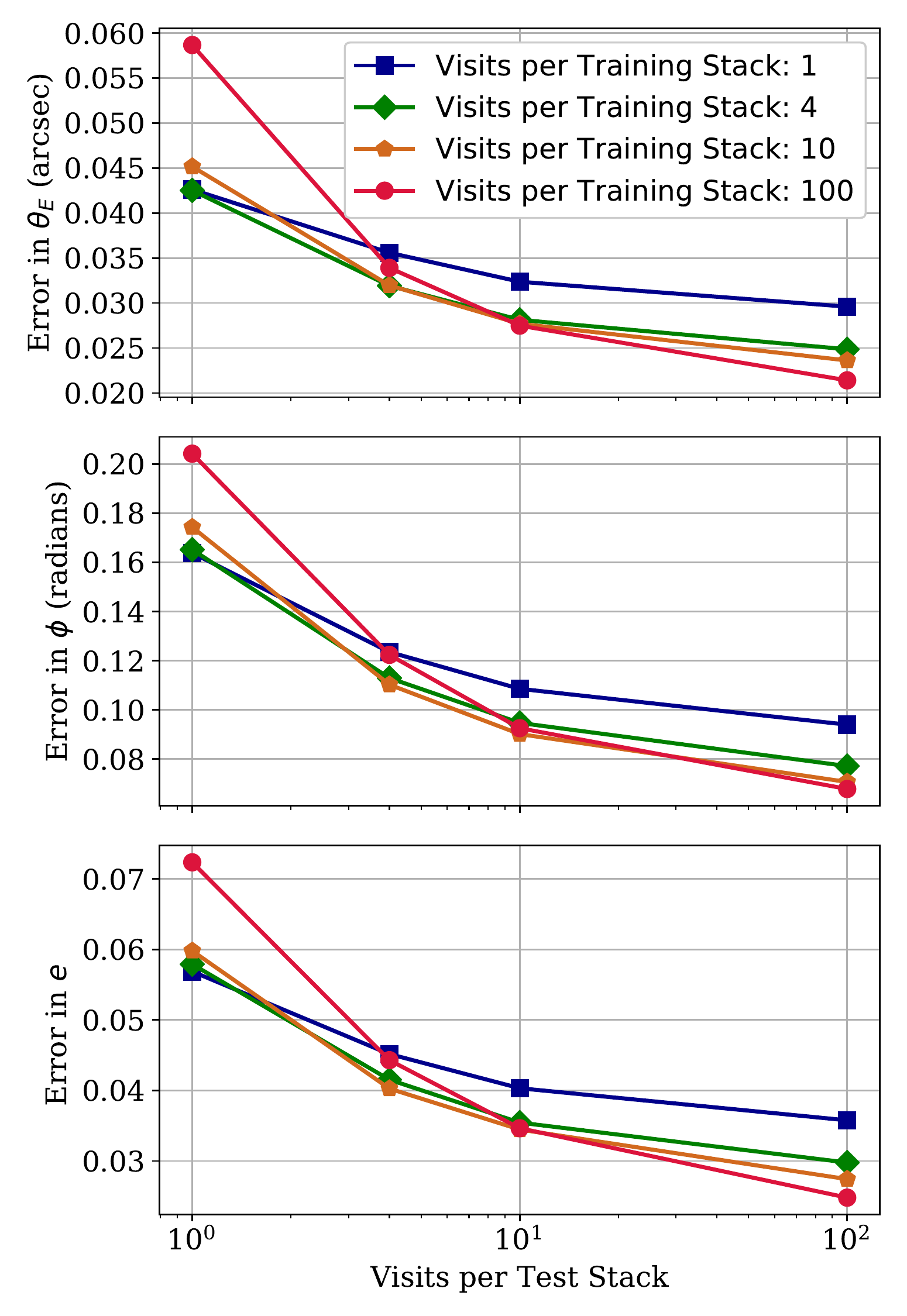}
     \caption{Lens model parameter uncertainties as a function of the number of visits making up each stacked image in the test set. From top to bottom are the Einstein radius, orientation and ellipticity of the lens mass profile. In each case, blue squares, green diamonds, orange pentagons and red circles correspond to where the network was trained on stacked images made up of 1, 4, 10 and 100 visits, respectively.}
    \label{fig:euclid_fn_of_stacks}
\end{figure}

The results of this are presented in Fig. \ref{fig:euclid_fn_of_stacks}, showing the uncertainties of the mass model parameters as functions of the number of visits making up stacked images in the test data set. The figure shows how network accuracy always improves when applied to more deeply stacked test images. For example, when trained on single-visit images, testing on stacked images of 100 visits rather than single-visit images results in an average decrease in uncertainty of $37\pm5$ per cent across the parameters \{$\theta_{E}$(arcsec), $\phi$(rad), $e$\} shown in the figure, from values of \{0.04, 0.16, 0.06\} to \{0.03, 0.11, 0.04\}. Furthermore, while the network accuracy improves as expected when tested on deeper stacks, there is a notable increase in accuracy when the network is retrained on multivisit stacks. Compared to training and testing on single-visit images, an average decrease in uncertainty of $55\pm4$ per cent is observed if instead trained and tested on stacked images of 100 visits, resulting in values of \{0.02, 0.07, 0.02\}.

The results also indicate that the fewer visits making up the training stacks, the better the network does when presented with single-visit and fewer-visit stacks. A network trained on single-visit images achieves uncertainties that are 1.2-1.4 times smaller than a network trained on 100-visit images when tested on single-visit data sets.

\section{Summary and Discussion} \label{sec:summary_discussion}

Strong gravitational lensing provides a means of studying the properties of distant galaxies, with upcoming surveys set to produce many thousands of images of such events. In preparation for this, a convolutional neural network was constructed to estimate the mass model parameters of the lenses from images. Simulated LSST and Euclid images were made for network training, utilizing SLACS lens data and modelling the lenses with SIE mass models. This resulted in the creation of strong galaxy-galaxy lens images characteristic of those produced by the Euclid VIS filter and LSST (for both $r$-band and multiband $gri$ imaging). The simulations allowed for the fast generation of guaranteed lenses, and benefited from the use of multiple SEDs for various galaxy ages and metallicities, along with realistic distributions for parameters such as axis ratios and mass-light profile scatter. The CNN configuration was optimized while remaining relatively simple, and the network was trained on 50,000 images to estimate three mass model parameters: Einstein radius, and two components of complex ellipticity later converted to ellipticity and orientation.

The key findings of this work are as follows:
\begin{itemize}
% - Inclusion vs removal of lens light
    \item Inclusion of the foreground lens light gives an acceptable accuracy for predicting mass parameters, achieving uncertainties on \{$\theta_{E}$(arcsec), $\phi$(rad), $e$\} of \{0.06, 0.23, 0.08\} for Euclid VIS, \{0.09, 0.22, 0.08\} for LSST $r$-band and \{0.07, 0.17, 0.06\} for LSST $gri$ images. When trained and tested on images with lens light removed, the CNN results in uncertainties of \{0.04, 0.16, 0.06\} for Euclid VIS, \{0.05, 0.15, 0.05\} for LSST $r$-band and \{0.04, 0.11, 0.04\} for LSST $gri$ images. Hence, removal of the lens light results in errors being reduced by 27-41 per cent (an average of $34\pm5$ per cent) depending on the predicted parameter and the data set, with the greatest improvements seen for Einstein radius and LSST $gri$ images.
% - Single-band vs multiband images
    \item Significant improvements to the accuracy of the network are obtained when using multiple bands, reducing the uncertainties on \{$\theta_{E}$(arcsec), $\phi$(rad), $e$\} from \{0.09, 0.22, 0.08\} to \{0.07, 0.17, 0.06\} (an average decrease of $20\pm2$ per cent) for images with lens light, and from \{0.05, 0.15, 0.05\} to \{0.04, 0.11, 0.04\} (an average decrease of $24\pm2$ per cent) for images with lens light removed. As such, the inclusion of colour information results in an accuracy comparable to single-band images without lens light.
% - Mass-light alignments when including lens light
    \item For data sets containing the foreground lens light, differences between the light and mass profile ellipticity and orientation have a significant impact on the performance of the network. This is especially true when the network is trained on data sets with no or little difference between the two, due to the network being more influenced by the lens light as opposed to the lensed source light. In addition, even when trained and tested on images with zero scatter between light and mass profiles, uncertainties are an average of 11 per cent larger than those for images with their lens light removed, achieving values for \{$\theta_{E}$(arcsec), $\phi$(rad), $e$\} of \{0.015, 0.032, 0.013\} with lens light and \{0.015, 0.027, 0.011\} without. Notably, no systematic biases were observed in the results, provided the network was trained with some degree of mass-light scatter. This provides reassurance in measuring $\rm H_0$ from gravitational time delays, as a lack of systematic error ensures accuracy can be improved simply through increasing the sample size of time delay lenses. As a general rule, if alignment between mass and light is unknown, it is better to train a network assuming a moderate amount of scatter in the mass-light orientation and ellipticity offset than to train assuming that mass closely follows light.
% - Signal-to-noise
    \item The error in predicted parameters decreases approximately exponentially with increasing test image SNR up to 100, after which there is little improvement, with SNR>400 images having uncertainties that are between 1.6 and 2.7 times smaller than those of SNR=20-30 images. When trained on single-visit Euclid images, the uncertainties on \{$\theta_{E}$(arcsec), $\phi$(rad), $e$\} are reduced from \{0.04, 0.16, 0.06\} to \{0.03, 0.11, 0.04\} (an average decrease of $37\pm5$ per cent) when tested on stacks of 100 images compared to testing on single-visit images. The errors are reduced more, by an average of $55\pm4$ per cent, if trained and tested on 100-visit stacked images. However, whilst network error is always lower when testing with higher SNR images, applying a network trained on high SNR images performs worse on low SNR test images than a network trained on low SNR images.
\end{itemize}

Our network has relatively few layers compared to other CNNs \citep[e.g.][]{hezaveh2017fast}, but this does not appear to be detrimental, with the accuracy instead limited by the resolution of the images. To improve the network for application to real data would involve more realistic simulations and significantly larger training data sets to account for a wider variety of potential images. The method used by \cite{hezaveh2017fast} to predict mass model parameters involved training a combination of four networks: Inception-v4 \citep{szegedy2017inception}, AlexNet \citep{krizhevsky2012imagenet}, Overfeat \citep{sermanet2013overfeat} and their own network. The larger number of layers used in these networks was required to train on higher resolution images which incorporated more complex features such as hot pixels, cosmic rays, pixel masking and occasional background objects.

One simplification we have made is regarding the removal of lens light. For data sets in which the lens light was removed, this removal was for the ideal case when the lens light ellipticity and orientation are exactly known, but in reality this would not necessarily be the case. Nevertheless, our removal of lens light does leave behind realistic shot noise residuals that having been trained on may alleviate some of the difficulty in testing the network on images with more realistic lens light-removal. Additionally, when dealing with real images, masks are often included to cover residual light for all but the lensed source. In this work, no masks were applied to the simulated images, so the network would need to be retrained to cope with such additions.

In this work, we have assumed a lens population of isolated elliptical galaxies, adequately described by an SIE mass density profile without external shear. In our test data sets, the velocity dispersions of the foreground lens population seen in Fig. \ref{fig:dist_vel_disp} may be higher on average compared to the distribution of real lenses \citep{collett2015population}. This would correspond to the real population containing fewer lenses with large Einstein radii, resulting in a slight increase in overall CNN error, as per the top row of Fig. \ref{fig:error_vs_parameters}. When creating our simulated lensed images, we have assumed that the lens light centroid is exactly aligned with the centre of mass (but that the lens is not exactly centred in each image). In reality, lenses can rarely be regarded as being completely isolated, instead being subjected to perturbative effects from environmental structures, substructure in the lens itself and mass along the line of sight. These simplifications have enabled a more efficient investigation of the practicalities of using CNNs for lens modelling but the problem of how well CNNs trained on simple models generalise to real data requires further work.

\section*{Acknowledgements}

We thank the anonymous referee for their helpful comments and suggestions.
We also greatly appreciate the support and contributions made by Tony and Alex Pearson.
NL acknowledges support by the UK Science and Technology Facilities Council (STFC).
SD is supported by a UK STFC Rutherford Fellowship.

%%%%%%%%%%%%%%%%%%%%%%%%%%%%%%%%%%%%%%%%%%%%%%%%%%


\bibliographystyle{mnras}
\bibliography{references}

%%%%%%%%%%%%%%%%%%%%%%%%%%%%%%%%%%%%%%%%%%%%%%%%%%

%%%%%%%%%%%%%%%%% APPENDICES %%%%%%%%%%%%%%%%%%%%%

%\appendix


%%%%%%%%%%%%%%%%%%%%%%%%%%%%%%%%%%%%%%%%%%%%%%%%%%

% Don't change these lines
\bsp	% typesetting comment
\label{lastpage}
\end{document}